# Design and characterization simulation of Ti: sapphire-based femtosecond laser system using Lab2 tools in the NI LabView

M Hussain [a], T Imran [b]

[a] GoLP, Instituto de Plasmas e Fusão Nuclear, Instituto Superior Técnico, Universidade de Lisboa, Av. Rovisco Pais, 1049-001 Lisbon, Portugal

[b] Department of Physics & Astronomy, College of Science, King Saud University, 11541, Riyadh, Saudi Arabia

E-mail address: mukhtar.hussain@tecnico.ulisboa.pt

**Abstract-** We report on the 825-nm center wavelength, 9.17 mJ pulse energy Ti:sapphire-based femtosecond laser system simulation carried out by Lab2 tools in LabVIEW (National Instruments, Inc.). The design investigation and characterization of stretched, amplified and compressed pulses made by intensity module and second harmonic generation (SHG) frequency-resolved optical gating (FROG) module in Lab2. The minimum pulse duration of ~37.80 fs at the output of the compressor end obtained by simulations. The variation of pulse energy, FWHM and central wavelength versus number of passes in the amplifier are computed. The lab2 tools help to design and characterize laser system before to set up on the optical table. The simulation results save time to calculate parameters which are essential in femtosecond laser system designing. The Lab2 simulation tools, along with financial constraints, it is easier, simple and efficient to obtain results in short time.

## 1. INTRODUCTION

The latest progress in the high-peak-power femtosecond laser systems [1-5] are relatively significant for certain experimental applications. Such as plasma, optical field ionization [6], white-light continuum generation (WLC) [7-9] and high harmonic generations (HHG) [10]. Through the harmonic upconversion [11-15], the X-rays and short ultraviolet pulses can be generated by employing femtosecond laser systems during the phenomena of scattering off electron rays [16-18] and laser induced plasma [19]. The ultrashort hard X-rays and soft X-rays can be employed to investigate the short and long-range dynamics of the atom and to observe the development of highly excited system [20]. Such laser can also employ for the generation of coherent X-rays with a pulse duration of 100 attoseconds [21-22]. The various diagnostic techniques are used to investigate and characterize the femtosecond laser pulses; these techniques include autocorrelation [23, 24], frequency-resolved optical gating (FROG) [25-31] and spectral phase interferometry for dielectric electric field reconstruction (SPIDER) [32]. There is few commercially available software [33-36] which are used in femtosecond laser system design, in contrast, Lab2 [37] is free, easier to use and more versatile especially for conventional femtosecond laser system study and design.

The present work dedicated to simulating the designing and characterization of the Ti:sapphire-based femtosecond laser system using Lab2 tools in LabVIEW. Before to set up the laser system on the optical table the Lab2 tools assist in the study and optimization of various design parameters efficiently in less time. In this simulation, we have studied the design and development of Ti:sapphire-based amplifier, grating-based stretcher, and compressor. In the lab2 tools, we have set different simulation parameters such as grating separation, grating rulings, the incidence gratings angle, the number of passes in amplifier, crystal length, crystal angle, and the pump fluence. The pulses of the simulated femtosecond laser characterized by using intensity module and second harmonic generation (SHG) frequency-resolved optical gating (FROG) module in lab2 tools. In fact, in FROG technique auto-correlation signal is spectrally resolved and quite efficient in characterizing the spectral and temporal profiles of the femtosecond laser. We have studied a full-width half maximum, of the spectra at different stages of the laser system modules in Lab2. By keeping the simulation in observation, the femtosecond laser system can be designed for best possible and optimized femtosecond energy per pulse and pulse duration.

## 2. DESIGN SIMULATION OF FEMTOSECOND LASER SYSTEM

In our simulation, we have adopted the familiar concept of chirped pulse amplification (CPA) for high energy pulses as shown in Figure 1 (a), where chirp introduced in the stretcher which stretches the pulse in the time domain, this stretched pulses amplified by directing to the multipass amplifier. The extracted pulses from the amplifier directed to the compressor which compensate the dispersion introduced by the stretcher, the output pulse further characterized and diagnosed by introducing a spectrometer, power meter, and SHG-FROG.

The mode-locked femtosecond pulses of Gaussian nature with a central wavelength of 825 nm, 1.0 nJ pulse energy with temporal FWHM of 30 fs are obtained from the Gaussian pulse module which acts as an oscillator (Figure 1 (b)). The femtosecond pulses from the oscillator stretched temporally by the grating stretcher module. The amount of the stretched pulse, which be contingent on the grating rulings, the angle of incidence of a pulse on the grating, dispersion of grating and the displacement '$g$'. The displacement '$g$' is needed to be adjusted in such way that the second order phase change should be minimum which leads the maximum peak intensity. The simulation shows that separation between the gratings is adjusted in a way to maximize the peak intensity of output pulse and minimize the second order phase distortion of the input pulse. The modulation of phase can be obtained as frequency function [36],

$$\emptyset(\omega) = \frac{\omega}{c_0}\left[8f + 2r - 4g\frac{cos\beta_0 - sin(\beta_0-\beta)sin\alpha}{cos\beta}\right] + \frac{8\pi g cos\beta_0}{d}\tan\beta \quad (1)$$

Where $\alpha$ is the incident angle of the pulse on grating, $d$ is the constant of grating, $\beta_0$ is the central wavelength diffraction angle and $\beta$ is known as diffraction angle which is defined as,

$$\beta = \sin^{-1}(\frac{2M\pi c_0}{\omega d} - sin\alpha) \quad (2)$$

(a)

(b)

**Figure 1.** (a) Conventional scheme of CPA system, (b) Lab2 scheme of femtosecond laser system

The stretched pulses directed to the multi-pass (8-pass) amplifier. The cluster of amplifier simulates the two-sided crystal pumped by a green laser (527 nm). The amplifying crystal (Ti: sapphire) is pumped at the Brewster angle rather than normal incidence to evade reflection losses. The characterization of the amplifier is based on its active medium length and ion concentrations. The variation of pulse energy, FWHM and central wavelength versus passes in the amplifier computed to check its efficiency.

The modulation in phase which introduced by the grating stretcher compensated by the compressor. The simulation shows that the separation between the gratings must be adjusted per pulse requirements to compensate the dispersion. In Lab2 simulation, the saturation duration during the phenomena of amplification, active medium dispersive effects, and resonant amplification dispersive effects are accounted by the module.

The simulated femtosecond laser pulses characterized in time and spectral domain by SHG-FROG measurements technique [23, 25]. The FROG signal which is a function of both time delay and frequency expressed as

$$F(\omega,\tau) \propto \left|\int_{-\infty}^{\infty} dt\, f(t,\tau)e^{-i\omega t}\right|^2 \quad (3)$$

Where, $f(t,\tau)$ function can be calculated by employing non-linear process. While the SHG-FROG [23, 25] traces can be related as

$$I_{FROG}(\omega,\tau) = \left|\int_{-\infty}^{+\infty} E(t)E(t-\tau)\exp(i\omega t)\,dt\right|^2 \quad (4)$$

## 3. SIMULATION RESULTS

In lab2 tools, we set seed laser pulse parameters based on conventional femtosecond laser system as shown in Figure 2. The 30 fs pulses with 1-mm beam diameter Gaussian seed laser pulse source, which stretched to 19.87 ps with an incident angle of $30.0^0$, stretcher grating having 600 lines/mm rulings, the focal length of mirror 100.0 cm. The grating displacement factor '$g$' is kept 16.0 cm while the grating mirror focal length set to 10.0 cm and the 20.0 cm size value of the grating in the stretcher used. The stretched pulses then directed to the 8-pass amplifier. In the 8-pass amplifier, 13.00 mm Ti: sapphire crystal which cut at Brewster angle, pumped at 527.0 nm for pump duration of 160.0 ns. The variation of pulse energy, FWHM and central wavelength versus number of a pass in the amplifier are computed shown in Figure 2 & 3.

The extracted pulses from the 8-pass amplifier are incident at $30.0^0$ to the 600.0 lines/mm rulings grating compressor. For minimum pulse duration, the delta (distance between gratings measured at center wavelength/frequency) set to 32.85 cm while the focal length of the grating mirror and grating size is chosen to 10.0 cm and 20.0 cm, respectively. The evaluation of energy per pass in the 8-pass amplifier is shown in Figure 2 (a), which depicts the linear increase in energy with the first 6-pass in the amplifier while in the remaining passes due to cavity losses do not show such behavior.

Figure 2. (a) Energy variation of amplifier, (b) Fluctuation in the central wavelength of amplifier, (c) FWHM variation in the 8-pass amplifier

The stability of central wavelength observed, variation in the central wavelength along the number of passes in the amplifier is computed as shown in Figure 2 (b), which depicts the minute variations in the central wavelength in the first five passes while in the subsequent passes central wavelength shift to ~820 nm. This variation in the wavelength of the amplifier is due to the group velocity dispersion (GVD). The variation in the FWHM of the pulse in the amplifier along the number of passes is computed which shows the sharp reduction in the FWHM in the first 6-pass from 31.25 nm to 28.25 nm. The minimum FWHM obtained after 6-pass which is 28.25 nm while in the remaining two passes it jumps to 30.10 nm as shown in Figure 2 (c).

The extracted pulses from the 8-pass amplifier are directed to the grating-based compressor incident at an angle of $30^0$. The compressed pulses at the compressor end have a pulse duration of 37.80 fs with the pulse energy of 9.17 mJ. The spectral behavior of the compressed pulse centered at 825 nm which measured by employing spectrometer module as shown in the Figure 3. A comparison between the different number of passes in the amplifier is carried out at the compressor end and its effect on the temporal profile and energy of the simulated pulses calculated as shown Fig 4. (a) & (b).

Figure 3. Spectral profile at the compressor end

The 8-pass amplifier which is optimized one with the lowest possible pulse duration (37.80 fs), reasonable high energy per pulse (9.17 mJ). It is clear that pulse energy and pulse duration are increased with the number of passes, so optimized pulse duration achieved after 8-pass with reasonable energy per pulse.

The SHG-FROG image is shown in Figure 5 (a). In the module of FROG, the twofold copies of femtosecond pulses mixed in non-linear medium and the spectrum engendered the non-linear medium is measured between twofold copies as time delay function. The data from the SHG-FROG traces extracted by using FROG software. The grid size of the FROG module taken as 256×256 with 6.0 fs time step. The SHG-FROG trace retrieved by using the commercial software [32] from the directly measured FROG trace. The evolution of temporal and spectral profiles of the pulses plotted from the retrieved FROG trace. The temporal and spectral FWHM ~37.80 fs and 29.65 nm calculated with low temporal phase variations about 1 radians peak-to-peak as shown in Figure 6 (a & b).

(a)

(b)

Figure 4. (a) FWHM variation with number of pass in amplifier at compressor end (b) Energy per pulse variation with number of pass in amplifier at compressor end

(a)

(b)

Figure 5. (a) SHG-FROG image (inset) and retrieved temporal profile (b) retrieved spectral profile

## 4. CONCLUSIONS

The simulation of Ti: sapphire-based chirped pulse amplification laser system carried out by Lab2 tools in NI LabVIEW which results in the pulse energy of ~9.17 mJ, the central wavelength of 825.0 nm and the pulse duration of ~37.80 fs is computed. The energy variation per pulse, central wavelength fluctuation and FWHM fluctuation with the number of passes in the amplifier computed and optimized. The intensity module, spectrometer module and SHG-FROG module employed for the optimization and characterization of pulses across amplifier and compressor ends. The retrieved temporal and spectral evolution of the SHG-FROG traces depicts the pulse duration of ~37.80 fs with flat temporal profile and phase variation of ~1 radian over the bandwidth of 29.65 nm. These conclude the design study of the conventional femtosecond laser system; we can implement these simulations results to setup the laser system on the optical table, further optimization can be done by the simulations obtained by FROG module in Lab2.

## 5. ACKNOWLEDGEMENTS

The authors acknowledged the financial support from the King Abdulaziz City for Science and Technology (KACST) under the project number MS-36-2 to carry out this work.

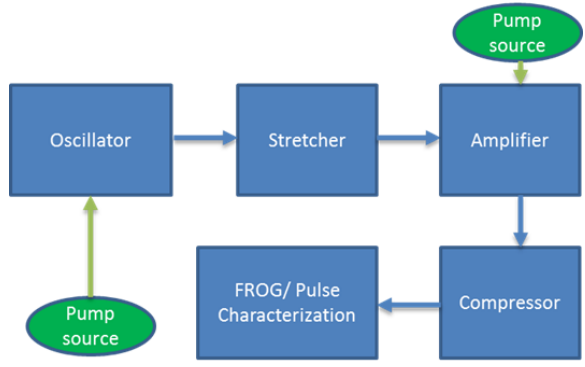

(a)

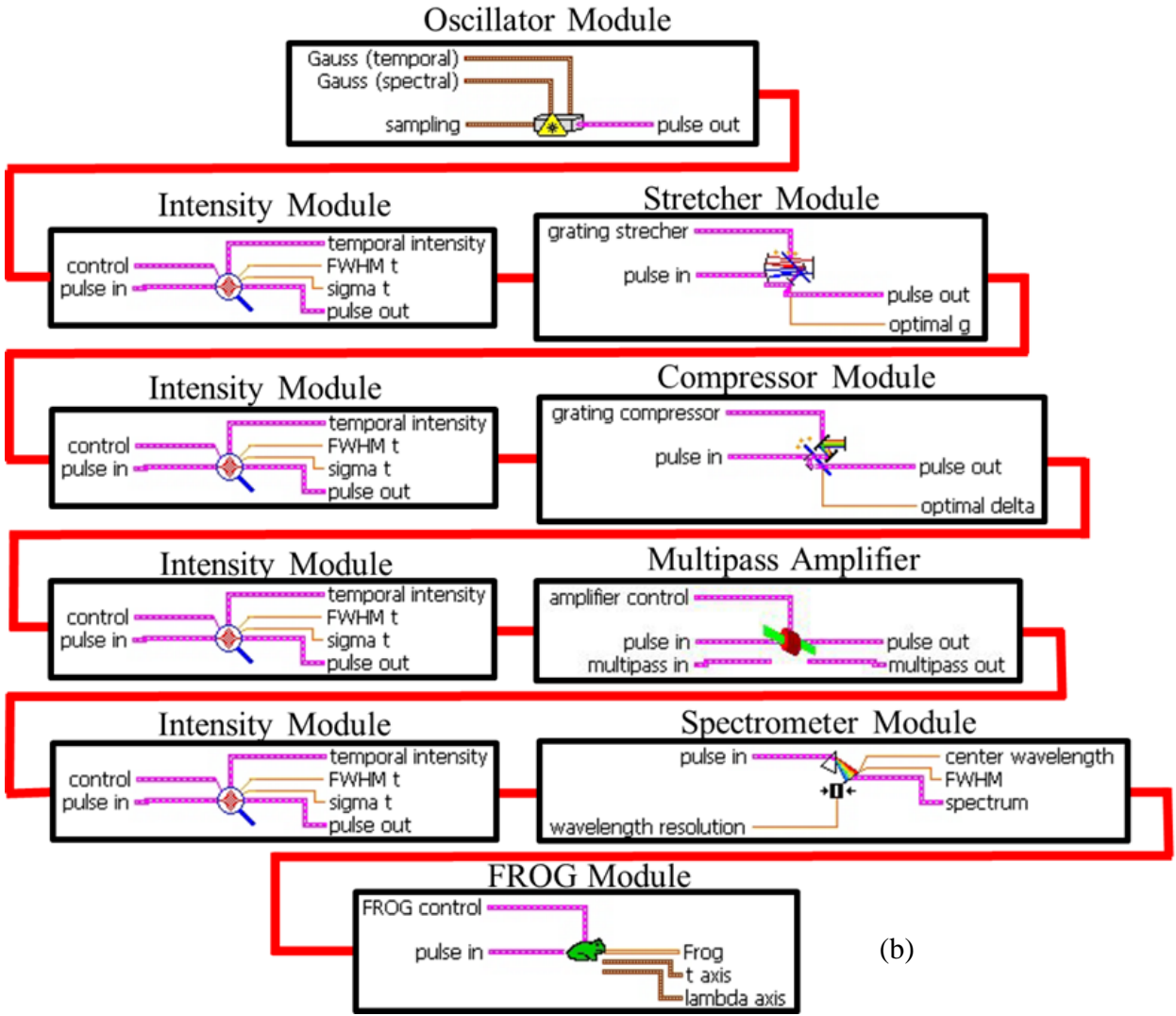

(b)

Figure 1 (a) Conventional scheme of CPA system, (b) Lab2 scheme of femtosecond laser system

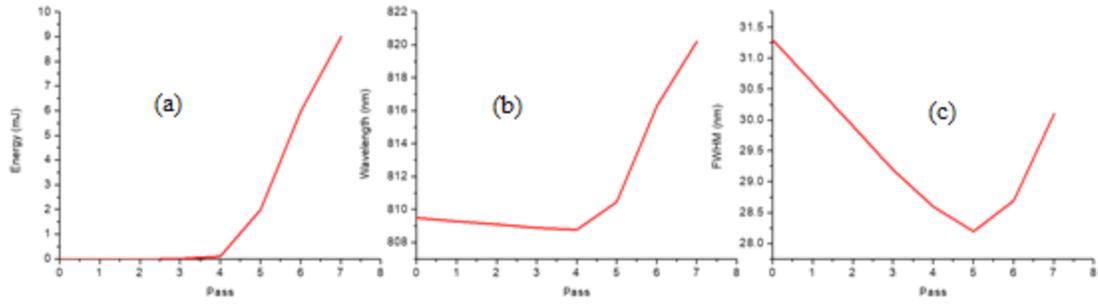

Figure 2. (a) Energy variation of amplifier, (b) Fluctuation in the central wavelength of amplifier, (c) FWHM variation in the 8-pass amplifier

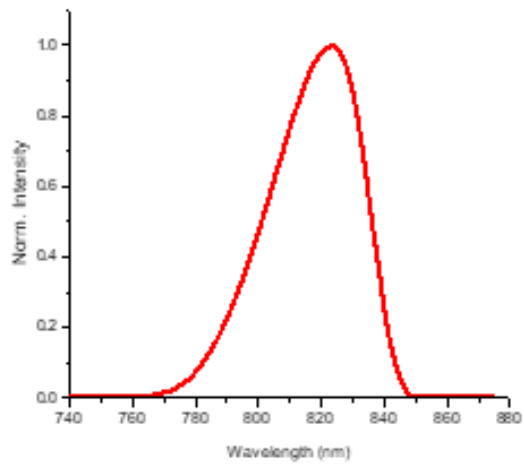

Figure 3. Spectral profile at the compressor end

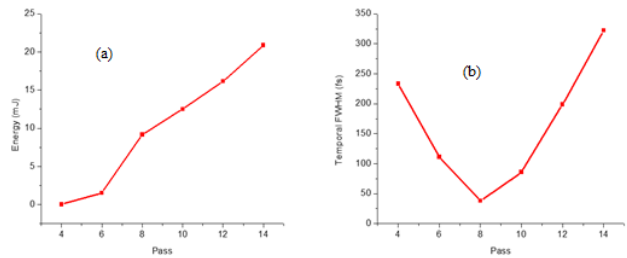

Figure 4. (a) FWHM variation with number of pass in amplifier at compressor end (b) Energy per pulse variation with number of pass in amplifier at compressor end

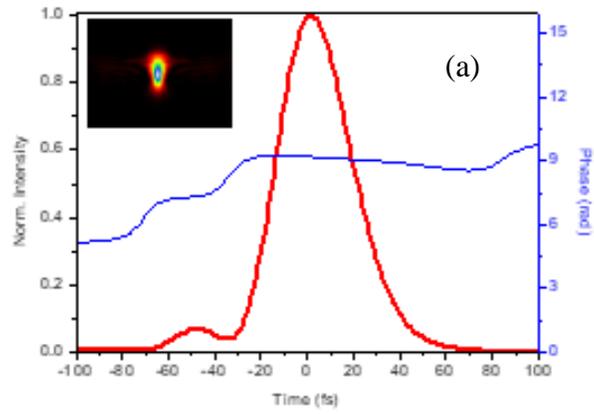

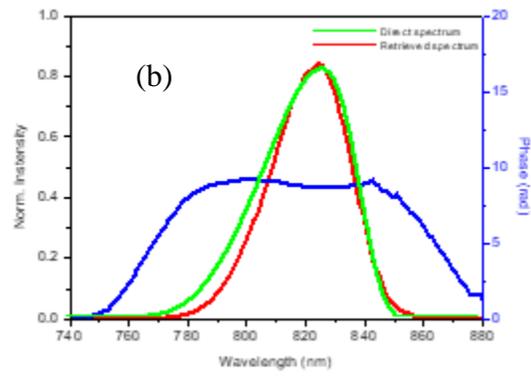

Figure 5 (a) SHG-FROG image (inset) and retrieved temporal profile (b) retrieved spectral profile